\documentclass{proclund}
\usepackage{color}

\newcommand{\beq}{\begin{equation}}
\newcommand{\be}{\begin{eqnarray}}
\newcommand{\eeq}{\end{equation}}
\newcommand{\ee}{\end{eqnarray}}
\newcommand{\ba}{\begin{array}}
\newcommand{\ea}{\end{array}} 
\newcommand{\bb}{\begin{thebibliography}}
\newcommand{\eb}{\end{thebibliography}}

\begin{document}
\pagestyle{prochead}


\title{{\color{red} DEUTERON STRUCTURE AT SMALL $N - N$ DISTANCES FROM
 INELASTIC $e-D$ AND $p-D$ REACTIONS}}
\author{\color{blue} {\underline{A.Yu.~Illarionov}}}
  \email{Alexei.Illarionov@jinr.ru}
  \homepage{http://thsun1.jinr.ru/~illar}
\author{\color{blue} G.I.~Lykasov}
  \email{lykasov@nusun.jinr.ru}
\affiliation
 {141980, JINR, Dubna, Moscow region, Russia\\~\\
}


\begin{abstract}
\begin{center}

\begin{minipage}{140mm}
The outline of the presentation is as follows:
\begin{itemize}
\item[I.~~]
  Nucleon Distribution in the Deuteron from $e-D$ and $p-D$ Processes.
\item[II.~]
  Quark Distribution in Deuteron from its Fragmentation to Pions
  and Deep Inelastic $e-D$ scattering.
\item[III.]
  Difference of These Two Distributions
  and Its Possible Understanding.
\item[IV.]
  Conclusions.
\end{itemize}
\end{minipage}

\end{center}
\end{abstract}
\maketitle
\setcounter{page}{1}

\section{Nucleon Distribution in the Deuteron from $e-D$ and $p-D$ Processes}

Over the past decades the study of the short range
structure of atomic nuclei attracts attention of
theorists and experimentalists. In the more
conventional picture, the basic degrees of freedom
are point-like non-relativistic nucleons. The
distribution of these nucleons in nuclei can be
calculated within the many-body approach by
introducing the phenomenological Hamiltonian 
fitted to the nucleon-nucleon scattering data and 
to the properties of the few-nucleon bound states. 
The nucleon distribution in a deuteron or the square 
of the deuteron wave function (DWF) magnitude
$n_D(k) ~=~ |\Psi_D(k)|^2$ can be obtained by solving the 
Schr${\ddot{\mbox{o}}}$dinger equation for DWF with different $N-N$ 
potentials. 

\begin{figure}[htb]
  \begin{center}
    \includegraphics[width=12cm]{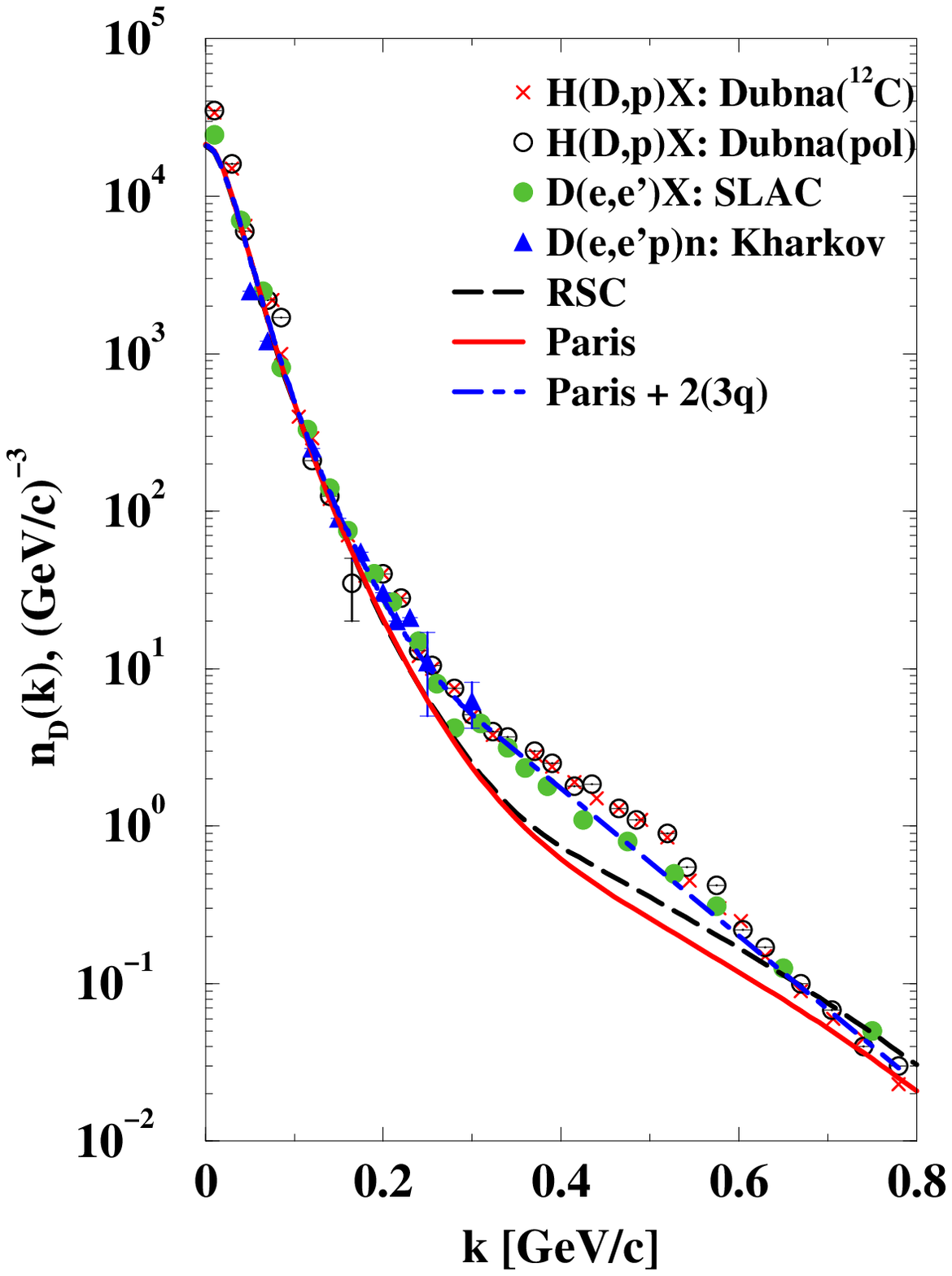}
    \caption{\label{Fig:WFD6q.eps}
 The nucleon distribution in the deuteron, $n_D(k) ~=~ |\Psi_D(k)|^2$
 extracted from the $e-D$ scattering and $Dp \to pX$ stripping data
 \cite{abl83}
 [{\bf \color{green}V.G.~Ableev {\it et al.}},
  Pis'ma Zh.Eksp.Teor.Fiz. {\bf 37}, 196 (1983)].
 The solid and dashed lines correspond to the calculations performed 
 within the impulse approximation. The dash-doted line show the effect
 of including of the non-nucleon component in the DWF
 [\cite{lyk93} {\bf \color{green}G.I.~Lykasov}, EPAN {\bf 24}, 59 (1993);
  \cite{efr88} {\bf \color{green}A.V.~Efremov {\it et al.}},
   Sov.J.Nucl.Phys. {\bf 47}, 868 (1988)].
      }
    \end{center}
\end{figure}

Experimentally $\Psi_D(k)$ is extracted usually from 
the elastic $e-D$ scattering and there is a
satisfactory theoretical description of these data
at small and moderate nucleon momenta $k$. This
information can be obtained also in experiments about the 
deuteron stripping $D p\rightarrow p X$ or semi-inclusive 
$e D\rightarrow e^\prime p X$ processes. The data about
$|\Psi_D(k)|^2$ obtained
from the experiments on $e-D$ scattering and 
$D p\rightarrow p X$ stripping are presented in Fig.~\ref{Fig:WFD6q.eps}.

The solid and dashed lines correspond to the calculations performed 
within the impulse approximation. As it has been shown
in \cite{lyk93}, the discrepancy between this lines
in Fig.~\ref{Fig:WFD6q.eps} and experimental points at 
$0.3$(GeV/c)$ < k < 0.45$(GeV/c) can be due to the secondary
interactions, namely the contribution of the so called
triangle graphs with the virtual exchange meson. 
Therefore this difference can not be interpreted as a 
contribution of the non nucleon degrees of freedom. 
However the high momentum tail of nucleon distribution
in a deuteron, e.g., $|\Psi_D(k)|^2$ at
$k > 0.65$(GeV/c) extracted from the experimental data
about $D p\rightarrow p X$ stripping \cite{abl83}
can not be described by using the standard degrees of
freedom in a deuteron, point-like non-relativistic
nucleons. There are models including a possible $2(3q)$ admixture
in the DWF, see for example \cite{bur84,kob82}, or non-nucleon 
degrees of freedom as like as $N N^*,\Delta\Delta$. etc.,
\cite{lyk93,efr88}, which allow to describe the experimental data about
$D p\rightarrow p X$ stripping at $k > 0.65$(GeV/c).

The Fig.~\ref{Fig:WFD6q.eps} shows the identity for $|\Psi_D(k)|^2$
extracted from $e-D$ scattering at not large $Q^2$ \cite{eD-data}
and $D p\rightarrow p X$ stripping data \cite{abl83}. However the
SLAC data allow to extract the nucleon distribution in a deuteron
at $k \leq 0.75$(GeV/c). Only experimental data about the
deuteron striping $D p \rightarrow p X$ \cite{abl83} result in an 
information about the high momentum tail of this distribution.

\section{Quark Distribution in a Deuteron from its Fragmentation to Pions}

The additional information about the deuteron structure at small
$N-N$ distances can be obtained from the analysis of the
fragmentation processes of a deuteron to pions emitted, mainly,
backward, e.g., $D p\rightarrow \pi X$ at large pion momenta 
or big values of the light-front variable
{\color{red}
           $$ z = \frac{p_\pi p_D}{p_p p_D} > 1. $$
}
where $p_\pi, p_D$ and $p_p$
are four-momenta of pion, moving initial deuteron and proton-target
at rest. For this type of the fragmentation process the inclusive
pion spectrum $\rho_{D p\rightarrow\pi X}(z)$ at large $z$ can give
us information about the valence quark distribution in a deuteron
$q_D(z)$ because at $z\rightarrow 2$:
{\color{red}
 $$
  \rho_{D p\rightarrow\pi X}(z) \propto q_D(z)~.
 $$
}

\begin{figure}[htb]
  \begin{center}
    \includegraphics[width=12cm]{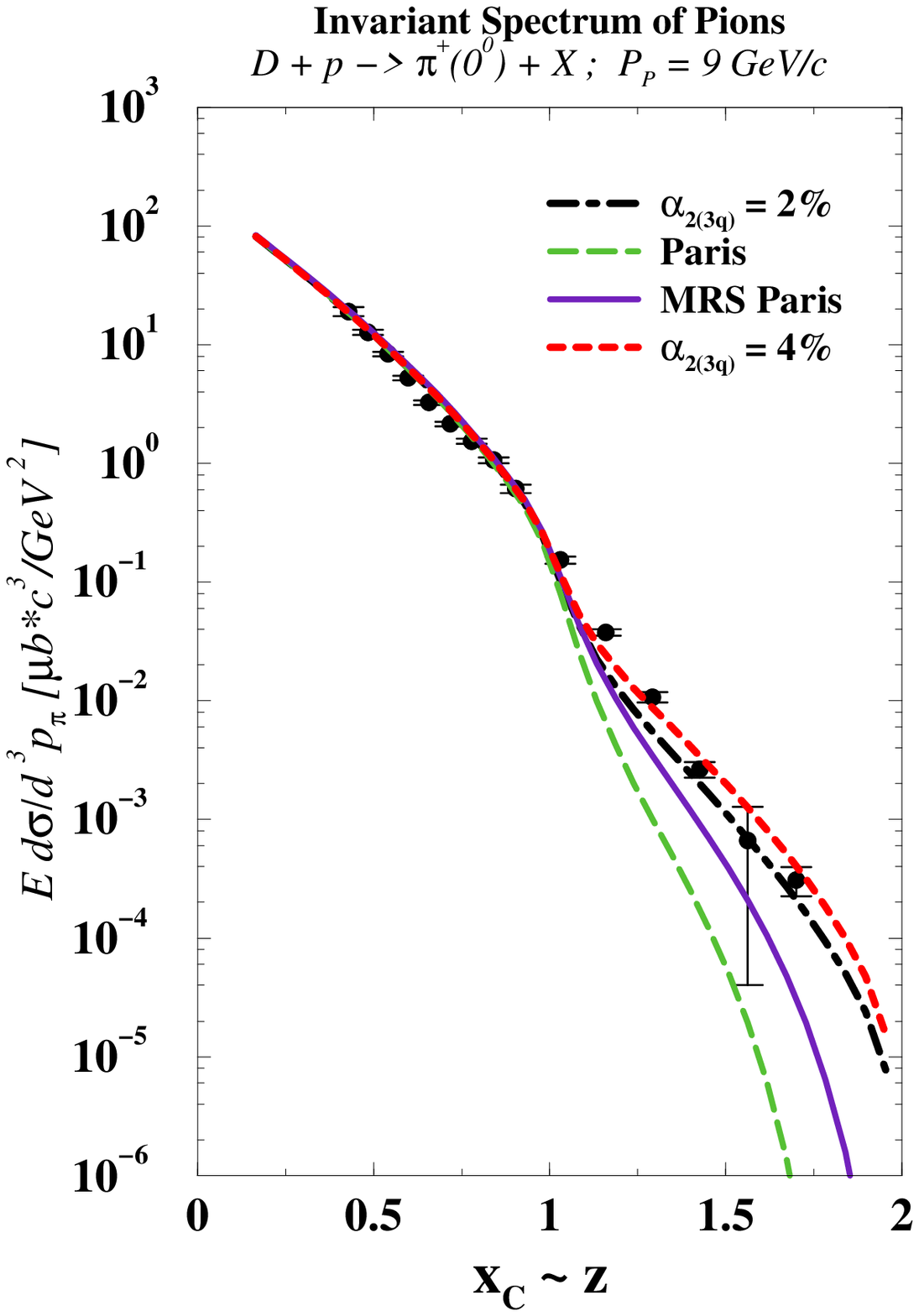}
    \caption{\label{Fig:pD.eps}
 The invariant spectrum of the backward pions in the deuteron
 fragmentation reaction via the cumulative variable \cite{sta79}
           $ {\rm x}_C \sim z = (\pi D)/(p D) > 1 $, 
 calculated in the relativistic impulse approximation
 with including of the non-nucleon component in the DWF
 [\cite{ill01} {\bf \color{green}A.Yu.~Illarionov {\it et al.}},
  Czech.J.Phys. {\bf 51} (2001), hep-ph/0007358, hep-ph/0012290];
 its probability {\color{red} $\alpha_{2(3q)}$ is $2\%$ - $4\%$}
 (dot-dashed and dashed curves, respectively). The long-dashed and
 solid lines correspond to the calculation using nonrelativistic DWF
 and DWF obtained by the minimal relativization scheme (MRS)
 [\cite{fs} {\bf \color{green}L.L.~Frankfurt and M.I.~Strikman},
   Phys.Rep. {\bf 160}, 236 (1988)], respectively.
 The calculated results are compared with the experimental data from 
 [\cite{bal85} {\bf \color{green}A.M.~Baldin}, Nucl.Phys. {\bf A434},
 695c (1985)] for a
 the projectile proton momentum of $P_p = 9$~GeV/c. One can see the good
 description of the experimental data for all cumulative $x_{\cal C}$
 \cite{sta79}.
      }
    \end{center}
\end{figure}

\begin{figure}[htb]
  \begin{center}
    \includegraphics[width=\columnwidth]{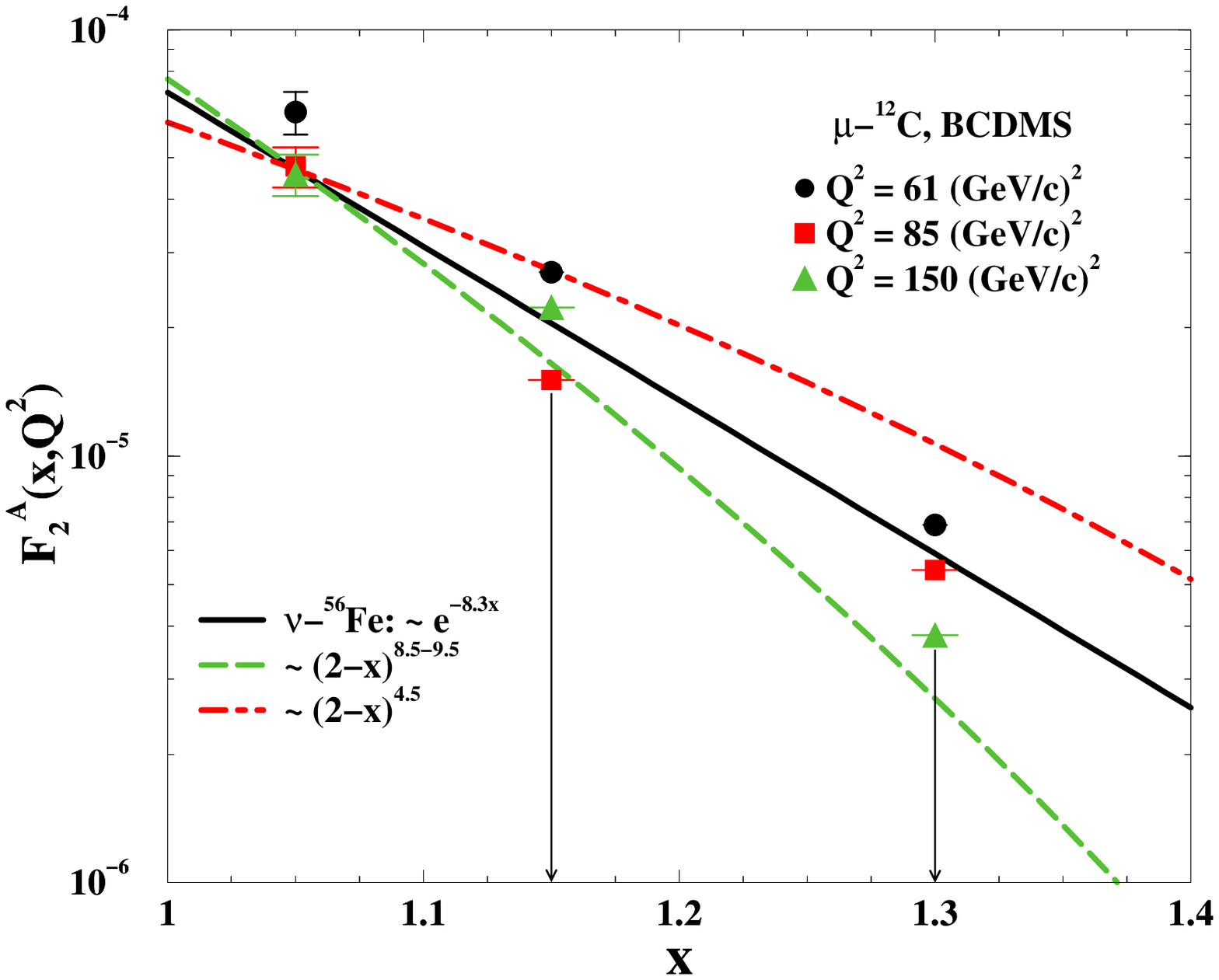}
    \caption{\label{Fig:F2.eps}
 This $x$-dependence of the structure function $F_2^A(x, Q^2)$ at $x > 1$
 from the deep-inelastic scattering (DIS).
 The experimental data are taken from
 \cite{BCDMS} ({\bf \color{green}BCDMS} muon DIS experiment) and
 \cite{CCFR} ({\bf \color{green}CCFR E770} neutrino DIS experiment).
      }
    \end{center}
\end{figure}

In Fig.~\ref{Fig:pD.eps} the pion inclusive spectrum
$\rho_{D p\rightarrow\pi X}(z)$ is presented.
This figure shows the main contribution to the
spectrum at $z > 1.6$ is coming from the high momentum tail of the
nucleon like objects in a deuteron $G_{2(3q)}$. This function 
$G_{2"N"}$ effectively includes the Fock columns $N
N^*,\Delta\Delta,\pi N N,...$ of the deuteron state
\cite{efr94,efr88} and \cite{ben97}. According to this approach the
valence quark distribution in a deuteron at $z > 1$, for example 
$u_D(z)$ has the following form:
{\color{red}
\begin{equation}
  u_D(z) = \frac{C_{uD}}{\sqrt{z}}(2 - z)^{4.5}~,
\label{def:uDz}
\end{equation}
}
where $C_{uD}$ is the normalized constant. On the other hand,
the quark counting rule results in at large $z$ the following
\cite{fs}:
{\color{red}
\begin{equation}
  u^{q.c.}_D(z) \sim \frac{1}{\sqrt{z}}(2 - z)^{10}~.
\label{def:uqq}
\end{equation}
}
Really, the $z$-behavior for $q_D(z)$ given by
Eq.(\ref{def:uqq}) is predicted by the perturbative QCD \cite{fs}.
This $z$-dependence of $q_D(z)$ or the deuteron structure function
$F_{2D}(z)$ at $z > 1$ has coincided with the measurements of 
the BCDMS \cite{BCDMS} and CCFR \cite{CCFR} (exp.~E770) collaborations,
Fig.~\ref{Fig:F2.eps}.

~\vspace{0.2cm}

{\color{blue} \centerline{\rule[1mm]{\textwidth}{1.0mm}}}
\vspace{0.5cm}

\centerline{\Large \bf \color{red} Q U E S T I O N}

\vspace{0.5cm}

Why is there a big difference between quark distribution in a deuteron at
large $z$ obtained from the deuteron fragmentation to pions and from deep
inelastic scattering?

\vspace{0.5cm}
{\color{blue} \centerline{\rule[1mm]{\textwidth}{1.0mm}}}
\vspace{0.5cm}

\begin{center}
It can be understood within the quark model suggested in \cite{ben97}.

%
\end{center}

\section{Difference of These Two Distributions and
         Its Possible Understanding}

According to \cite{ben97}, one can construct the quark distribution in
a deuteron at $z > 1$ calculating the relativistic invariant phase
space volume available to a quark when the quark distributions
of different nucleons overlap. Let us shortly present a scheme
of this approach. 
The quark distribution in a nucleon can be written in the 
following form:
{\color{red}
\begin{equation}
 q_N(z) \sim z^{a_N} (1 - z)^{b_N}~,
\label{def:qnz}
\end{equation}
}
where $a_N, b_N$ are some values which will be discussed little bit
later.
Calculating the overlap of the quark distributions of different
nucleons, according to \cite{ben97} one can get the quark distribution
in the overlapping region which at $z > 1$ coincides with 
$q_D(z)$,
{\color{red}
\begin{equation}
 q_D(z) \sim z^{a_N} (2 - z)^{2b_N + 2 + a_N}~.
\label{def:qdz}
\end{equation}
}
In principle, the $z$-dependence of distributions of the
constituent quarks in nucleon participating at soft hadron
processes is different from distributions of current valence 
quarks interacting with lepton beam in deep inelastic 
lepton-nucleon scattering (DIS). This difference is determined 
mainly by the value of $b_N$. So, for current quarks 
$b_N \simeq 3.5\div4.5$ at $Q^2 = Q_0^2 \simeq 2.\div3.$(GeV/c)$^2$,
according to experimental data on DIS, but the constituent quarks
have to have the true Regge asymptotic at large $z$ \cite{kai82} and
for them $b_N = \alpha_R(0) - 2\alpha_B(0) = 3/2$; here $\alpha_R(0)$
and $\alpha_B(0)$ are the intercepts of meson and average baryon 
Regge trajectory $\alpha_B(0) = -0.5$ and $\alpha_R(0) = 1/2$.
The value of $a_N$ determines the quark distribution in a nucleon
at $z \rightarrow 0$ and $a_N = -\alpha_R(0) = -0.5$ for the
constituent and current quark except the region of too small $x$
corresponding to last HERMES data.

Therefore inputting the different values of $b_N$ to the 
Eq.(\ref{def:qdz}) we get the distribution for constituent quarks
in a deuteron at $z > 1$:
{\color{red}
\begin{equation}
 q^{cons}_D(z) \sim \frac{1}{\sqrt{z}} (2 - z)^{4.5}
\label{def;const}
\end{equation}
}
and current valence quarks in a deuteron at $z > 1$:
{\color{red}
\begin{equation}
 q^{cur}_D(z) \sim \frac{1}{\sqrt{z}}(2 - z)^{8.5\div9.5}
\label{def:qcur}
\end{equation}
}
Actually, the $Q^2$ evolution of the current quark distributions 
in a nucleon using the Altarelli-Parisi-Dokshicer-Gribov-Lipatov (DGLAP)
equation \cite{gl72,ap77} has to be included by more careful
calculation of $q^{cur}_D(z)$.

\section{Conclusions}

\begin{itemize}
\item[I.~~]
One can extract the nucleon distribution in a deuteron over
momentum $k$ from experimental data about $e-D$ scattering
and $D p \rightarrow p X$ stripping within the impulse approximation
at small $k < 0.25$(GeV/c) only.
\item[II.~]
This procedure is incorrect at larger momenta up to
$0.5$(GeV/c) because the secondary graphs have to be included.
\item[III.]
The conventional deuteron wave function of type, for example
the Paris one, doesn't describe the high momentum tail of
protons in the deuteron stripping $D p \rightarrow p X$ at
$k > 0.5$(GeV/c).
\item[IV.]
Additional information about the deuteron structure at small
$N-N$ distances can be obtained from the deuteron fragmentation to pions.
\item[V.~]
The quark distribution in a deuteron describing the inclusive
pion spectrum in the process $D p \rightarrow \pi X$ at $z > 1$
is different from the analogous distribution describing the
DIS $e-D$ scattering at $x = Q^2/(2m\nu) >1$.
\item[VI.]
This difference can be understood within the approach which
is based on the calculation of the relativistic invariant
phase space volume available to a quark when the quark
distributions of different nucleons overlap.
\item[VII.]
The constituent quarks participate at the deuteron
fragmentation to protons or pions, on the other hand lepton
interacts with current quarks by the DIS $e-D$ scattering.
They have completely different $x$-distributions in a nucleon
especially at large $x$. It leads to the different distributions
of these sorts of quarks in a deuteron.
\end{itemize}

\section*{Acknowledgments}
This work has been supported by RFFI grants N 99-02-17727.

\end{document}